\documentclass[12pt,english]{article}
\usepackage[T1]{fontenc}
\usepackage[latin1]{inputenc}
\usepackage{graphicx}

\makeatletter

 \newcommand{\lyxaddress}[1]{
   \par {\raggedright #1 
   \vspace{1.4em}
   \noindent\par}
 }

\usepackage{babel}
\makeatother
\begin{document}

\title{\textbf{\Large Non-linear model equation for three-dimensional Bunsen
flames}}

\maketitle
\begin{center}Bruno Denet\end{center}

\lyxaddress{\begin{center}IRPHE 49 rue Joliot Curie BP 146 Technopole de Chateau
Gombert 13384 Marseille Cedex 13 France\end{center}}

\begin{center}submitted to Physics of Fluids\end{center}

\begin{abstract}
The non linear description of laminar premixed flames has been very
successful, because of the existence of model equations describing
the dynamics of these flames. The Michelson Sivashinsky equation is
the most well known of these equations, and has been used in different
geometries, including three-dimensional quasi-planar and spherical
flames. Another interesting model, usually known as the Frankel equation,
which could in principle take into account large deviations of the
flame front, has been used for the moment only for two-dimensional
expanding and Bunsen flames. We report here for the first time numerical
solutions of this equation for three-dimensional flames.



\newpage
\end{abstract}

\section{Introduction \label{sec:Introduction}}

The Michelson Sivashinsky equation, or simply Sivashinsky equation,
depending on the author \cite{siva77}, is a good example of a non
linear model equation that has obtained results far beyond any expectations.
While originally derived for flames with a low heat release (which
are rather rare indeed, since in typical flames the burnt gases have
a temperature five times higher than the fresh gases) this equation
has succeeded in providing a qualitative description of the dynamics,
even for large heat release. This equation has been used for fronts
in different geometries (planar on the average \cite{michelsonsiva},
expanding flames\cite{filyandsiva}, oblique flames \cite{bouryjoulin.ctm2002}),
but with always the limitation that the deviation (actually the variation
of the slope) from the unperturbed geometry must be relatively small,
and that the front has to be described by a function (of the lateral
coordinate, of an angle ...) 

Another possible description \cite{frankel} has been proposed in
1990 by Frankel and does not suffer from these drawbacks, i.e. its
formulation is coordinate-free, with no privileged direction. It was
pointed out to the author by one referee that an approach similar
to Frankel's (using a potential approximation for the flame generated
flow field) had already been used in a 1982 article by Ghoniem, Chorin
and Oppenheim \cite{ghoniemchorinoppenheim}, which could justify
changing the name of the equation (usually called Frankel equation).
However, contrary to \cite{frankel}, the front was not described
in a lagrangian way by a set of markers, but was calculated using
a SLIC method \cite{slic}. Another difference is that turbulence
was described in \cite{ghoniemchorinoppenheim} with vortex methods.
The two approaches are thus not completely equivalent. However, it
is true that several aspects of the work of Frankel were used by other
authors before the 1990 paper (see for instance \cite{pinderatalbot86}\cite{ashurst87}).

In \cite{frankel}, the flow is supposed potential everywhere, fresh
and burnt gases, and the problem is transformed into an integro-differential
equation involving Green's functions and the position of the front.
This may at first seem completely ridiculous; after all, it is known
that for large heat release, vorticity is generated at the front and
has to be present in the burnt gases, even if the flow is potential
in the fresh gases. But actually, this approximation is close to the
low heat release limit of the Sivashinsky equation, and it has been
shown in the 1990 Frankel article (a very important point of this
paper) that this equation reduces to the Sivashinsky equation for
plane on the average and circular flames. Surprising as it may seem,
the success of the Sivashinsky equation is actually a (qualitative)
success of the potential approximation. However, contrary to the Sivashinsky
equation, the Frankel equation has not been for the moment derived
rigorously from an expansion in powers of the density ratio starting
from the hydrodynamic equations. Higher order expansions in power
of the density ratio have lead in the plane case to the derivation
of extensions of the Sivashinsky equation \cite{kazakovpof2002} (see
also \cite{boury}). An alternative method has been proposed \cite{bychphysrep}
, based on a weak-nonlinearity approximation. This approximation has
been criticized in \cite{kazakovpof2002}.

The good qualitative agreement of the potential model with experiments
(see for instance the comparison in \cite{denetbunsen2d} or the figures
of the current article) is thus less surprising: the same behavior
is seen in the related Sivashinsky equation case. Naturally, the author
does not claim that the agreement is quantitative, as the potential
approximation is well-known to change the linear growth rate of the
Landau instability \cite{sivaclavin}. For large heat release, the
potential model does not satisfy the correct conservation laws at
the front, particularly the pressure jump condition, which explains
this discrepancy.

In this article, however, we will limit ourselves to the potential
model. Numerical solutions of this equation for two dimensional flames
are much more difficult than in the Michelson Sivashinsky case, where
typically periodic boundary conditions and fast Fourier transforms
are used. In the coordinate-free case, the front, which is a line,
is discretized with different marker points, which move because of
the fluid velocity and the flame speed. Nevertheless, simulations
of this equation have appeared in the expanding flame \cite{frankelsiva}\cite{blinnikovsasorov}
\cite{ashurst97}\cite{denetfrankel} and the Bunsen burner case\cite{denetbunsen2d}.
But for the moment no numerical solution of this equation has been
obtained for three dimensional flames, where the front has to be described
by a surface and not a line. On the contrary, the Michelson Sivashinsky
equation has been solved numerically for these flames (see for instance
planar flame solutions without and with gravity \cite{michelsonsiva}
\cite{denetnonlinear} and a recent paper on spherical flames\cite{dangelojoulinboury})

In this article we will show the first numerical solutions of the
Frankel equation for three dimensional flames, in the Bunsen burner
geometry. The equation will be presented, along with some technical
details of the numerical solution. Then results for flames with different
forcings and for polyhedral flames will be shown.

\section{Model and numerical method\label{sec:Model}}

Let us first recall some notations. The flame velocity relative to
premixed gases will be noted $u_{l}$( and will have the constant
value $1$ in all the simulations presented in this article) . If
we define $\rho_{u}$ the density of fresh gases, $\rho_{b}$ the
density of burnt gases, $\gamma=\frac{\rho_{u}-\rho_{b}}{\rho_{u}}$
is a parameter measuring gas expansion ($\gamma=0$ without exothermic
reactions, $\gamma$ being close to $1$ for large gas expansion).
This difference in density between fresh and burnt gases is the main
cause of the Darrieus-Landau instability of premixed flames. We use
in this article an equation relative to fresh gases, in the Bunsen
burner geometry, as in \cite{denetbunsen2d}, contrary to the original
article \cite{frankel}, where the equation was written relative to
burnt gases for the case of an expanding flame. The unburnt gases
are injected at the velocity $U$. We will consider in Section \ref{sec:Forcing}
that $U$ is constant in space and that the flame is attached on a
circle (i.e. we define points on this circle that do not move during
the simulation). In Section \ref{sec:Polyhedral}, on the contrary,
we shall see that in order to obtain the polyhedral flames observed
experimentally, a boundary layer has to be considered. 

$\kappa$ is the mean curvature at a given point on the flame , $\varepsilon$
is a constant number proportional to the Markstein length, $\overrightarrow{n}$
is the normal vector at the current point on the front, in the direction
of propagation. After some calculations similar to \cite{frankel},
an evolution equation is obtained, valid for an arbitrary shape of
the front, which is here a surface, contrary to \cite{denetbunsen2d},
where it was only a line.

\begin{equation}
V(\overrightarrow{r},t)=u_{l}+\varepsilon\kappa+\overrightarrow{U}.\overrightarrow{n}+\frac{1}{2}\frac{\gamma}{1-\gamma}u_{l}-\frac{u_{l}}{4\pi}\frac{\gamma}{1-\gamma}\int_{S}\frac{\left(\overrightarrow{\xi}-\overrightarrow{r}\right).\overrightarrow{n}}{\left|\overrightarrow{\xi}-\overrightarrow{r}\right|^{3}}dS_{\xi}+\overrightarrow{V}_{boundary}.\overrightarrow{n}\label{mathed:V}\end{equation}

This equation gives the value of the normal velocity $V$ on the front
as a sum of several terms, the laminar flame velocity with curvature
corrections, the velocity of the incoming velocity field and an induced
velocity field (all the terms where $\gamma$ appears) which contains
an integral over the whole shape (indicated by the subscript $S$
in the integral ). This integral is a sum of electrostatic potentials. 

The term $V_{boundary}$ is a potential velocity field (continuous
across the flame) added to the equation in order to satisfy the boundary
conditions. Here the condition is simply that

\[
\left(\overrightarrow{V}_{induced}+\overrightarrow{V}_{boundary}\right).\overrightarrow{n}=0\]

at the injection location, where $\overrightarrow{n}$ is parallel
to $\overrightarrow{U}$ , so that $V_{boundary}$ is given by the
same type of integral as $V_{induced}$ , but over the image of the
front (see \cite{denetbunsen2d}), which is a front symmetric of the
real flame in the symmetry $z$$\rightarrow-z$ (actually we have
chosen $z$$\rightarrow$ $-z-0.01$ in order to prevent problems
from occurring when a front crosses its image)

When the velocity $V(\overrightarrow{r},t)$ is obtained from equation
(\ref{mathed:V}), the marker points which define the surface are
moved :

\[
\frac{d\overrightarrow{r}}{dt}=V\overrightarrow{n}\]

where $\overrightarrow{r}$ denotes the position of the current point
of the front. 

The different marker points of the surface are linked by triangles.
A LGPL C library, gts \cite{gts} written by Stéphane Popinet, has
been used to describe the triangulated surface, perform the integrals
in equation (\ref{mathed:V}), and calculate the mean curvature and
the normal at any given point. The calculation of the curvature is
done by a method proposed in \cite{meyercurvature}. Let us note that
gts itself is implemented using glib, a well-known general purpose
library that is also used directly from the author's program in order
to manipulate lists and hash tables. A zoom of a surface shown later
(from a different angle) in Figure \ref{fig:blanc} (right part of
the figure) can be seen in Figure \ref{fig:zoom}. As we use a large
number of points, typically $20000$, the different points cannot
be seen if the whole mesh is shown.

Problems typically encountered in Lagrangian descriptions of two-dimensional
flames as discontinuities \cite{denetlagrange} \cite{lamtalbot2003}
are also present here. First of all, there is a need to adapt the
mesh, adding and removing points where necessary. As a result, the
type of algorithm used here is inherently noisy, and the noise is
higher when there is a rapid time variation. On the other hand, the
spatial precision of these Lagrangian descriptions is generally higher
than for Eulerian methods, but there is the problem of self-intersections,
which occur very often and must be accounted for. One of the interest
of this article is to show that it can be done even for three-dimensional
flames.

As in \cite{denetlagrange}, we perform the reconnections after the
intersection has taken place: we first detect the triangles that self-intersect
(this is actually a gts function, which is performed in an efficient
way), then these triangles are removed from the surface (except on
the boundary). We also remove triangles which are not connected anymore
to the main surface after the first triangle removals. At this stage,
the surface has one or more holes. We detect the contours of the different
holes, and construct small surface patches which are added to the
main surface. At this stage, the surface has no hole left, but the
patches can have normals oriented in the wrong direction. In order
to get a common orientation for the whole surface, we traverse the
surface starting from the boundary, looking in a recursive way at
each neighbor of a given triangle in order to ensure that the orientation
of the whole surface is compatible. 

In the simulations that will be presented, we remove pockets created
during self-intersections, but actually it would have been very simple
to keep pockets with a sufficient number of points. We have programmed
a function splitting a surface in its connex components, which are
then sorted according to their number of points. Instead of keeping
only the main surface (with the largest number of points), it would
have been only a matter of two lines of code to apply a different
criterion. But the previous experience of the author with two-dimensional
flames described in a Lagrangian way has shown that the formation
of pockets is generally not very important, except occasionally and
for very special forcings. But it is important to discard very small
pockets, which can produce all kind of disagreements: further (absolutely
unrealistic) splitting of small pockets which forces to reduce the
time step, and global inversion of the normal of a small pocket, which
actually would have collapsed had the time step been smaller. For
real flames naturally, the flame speed is higher when two fronts are
very close, which let these small pockets disappear quickly. 

Another problem of these simulations, once the intersections are under
control, is that the calculations of the integral in equation (\ref{mathed:V})
is very expensive, if done exactly. The problem is exactly an N-body
problem (coulombian interaction), where an exact solution for $N$
points, is proportional to $N²$. As has been already done in a combustion
context \cite{blinnikovsasorov}, it is possible to use approximate
algorithms which reduce the number of operations. The author himself
had always used the exact algorithm when solving the potential model
for two-dimensional flames, but the extra dimension has forced him
to look at alternatives. There are two main approximate algorithms
for the N-body problem. The Greengard and Rokhlyn algorithm \cite{greengardrokhlyn}
(fast multipole expansion) approximates the field by a multipole expansion,
the algorithm is $O(N)$ but the prefactor is very high. We have preferred
to use the Barnes Hut algorithm \cite{barneshut} which is $O\left(N\log N\right)$.
The idea of this algorithm is to represent the field at a given point
by the action of a list of masses, where masses sufficiently far are
concatenated in order to provide a good approximation of the field,
the masses at the different levels being organized in a tree. The
source code of this algorithm can be found in Joshua Barnes' web site
\cite{barnesweb}.

\section{Flames in the presence of forcing \label{sec:Forcing}}

We consider here Bunsen burner flames, with position of the flame
imposed on the boundary, i.e. the boundary point must be on a circle
of radius $R$ (in all the simulations, the value $R$$=$$0.5$ will
be chosen) with a value $z$$=$$0$ ($z$ is the vertical coordinate).
Other boundary conditions will be considered in section \ref{sec:Polyhedral}.
In a first step, we would like to show that the three-dimensional
Frankel equation contains effects qualitatively similar to those observed
with the two-dimensional version in the same geometry \cite{denetbunsen2d}.
The last article contained evidence of the amplification and saturation
of cells created by the Darrieus-Landau instability, which were convected
toward the tip of the flame by the incoming flow. Furthermore, a repulsion
effect of the side of the flame where forcing was imposed on the other
side was observed in the simulations. This effect was very close to
the experimental observations of \cite{searbytruffautjoulin}. Here,
instead of having a two-dimensional geometry, i.e. a very long rectangular
Bunsen flame where the forcing is the same on the long side, we have
the usual round burner Bunsen flame.

Let us impose a localized forcing on the flame. We take a sinusoidal
forcing, where a term $2\sin\left(\omega t\right)$ is added to the
normal velocity, in a small window $z\in\left[\,0.4,\,0.45\right]$
$\theta\,\in\left[\,0,\,\pi/10\right]$, $\theta$ being the polar
angle. We take the value $\omega\,=\,200$ which is rather high, but
allows to form several wavelengths on the flame. The other parameters
are $U\,=\,12$, $\gamma\,=\,0.85$, $\varepsilon\,=\,0.1$. The last
parameter and the radius will have the same value in all the simulations
presented. The time step is $\Delta t\,=\,5\,10^{-5}$ and will keep
the same value in all the simulations. This value of the time step
is chosen to keep the self-intersections occurring on the front simple.
If the time step is too high, very large self-intersections can occur
at each time step close to the tip and the precision of the lagrangian
description becomes low. We obtain a situation presented in Figure
\ref{fig:sinus}. This figure contains three flames. Actually, we
had some difficulties in producing different figures of flames with
the same scale, so we have included several flames in the same figures.
This also makes comparisons easier. The two flames on the left are
the same, but seen from a different angle, and correspond to a flame
soon after the perturbations produced by the Darrieus-Landau instability
reach the tip. If one looks first at the middle image, it can be seen
that the sinusoidal forcing is very localized and produces a small
perturbation close to the base of the Bunsen flame. But this perturbation
develops, is amplified because of the Landau instability, is convected
toward the tip and extends laterally, which was of course not observed
in two-dimensional experiments and simulations. It is also to be noted
that the cells produced by the instability, once sufficiently developed,
have a geometry very close to the hexagonal cells well-known in the
planar on average configuration. However, when the cell is near the
tip, this hexagonal shape becomes less apparent and the cell can be
closer from a rectangle.

If one looks now at the side view of the same flame (left of Figure
\ref{fig:sinus}), the repulsion effect of the side submitted to perturbations
on the other seen in two-dimensional configurations is also present
here. As a consequence, there is a slow deflection of the unperturbed
side toward the right. This repulsion effect was explained qualitatively
by an electrostatic analogy in \cite{denetbunsen2d} (let us remember
that the model used here is potential). But here this deflection seems
smaller, and is seen clearly only when the lateral extension of the
perturbations is sufficiently large. Actually, the two-dimensional
situation corresponds to a lateral extent infinite in the transverse
direction, which explains why in the three-dimensional case, a large
lateral extent is needed to obtain the same effect.

The third flame image (right of Figure \ref{fig:sinus}), is a flame
with the same physical parameters, but some time later. This flame
is higher (the height of the flame fluctuates typically in this way
during the temporal evolution). But the main effect is that two cells
are merging close to the tip. Apparently, the way we have imposed
the forcing does not produce absolutely regular cells, possibly because
the forcing zone is too extended. It has happened that one cell had
a smaller amplitude close to the base of the flame, and during its
translation toward the tip, its amplitude and longitudinal extension
have slowly reduced, and now clearly this cell (second cell from the
tip, which is very small) is being captured by the third cell, which
is very large. This cell merging is absolutely typical of the Landau
instability and is well known in the planar configuration, where it
helps produce the one-cell flame obtained for every domain size in
the Michelson-Sivashinsky equation. Actually, for very large flame
height, self-similar mergings have been predicted theoretically \cite{boury}
\cite{bouryjoulin.ctm2002}. 

After this study of the effect of a sinusoidal forcing on the flame,
we are now interested in the effect of a white noise forcing applied
at the base of the flame (sufficiently far from the boundary to prevent
problems). To be specific, we have added to the normal velocity at
points where $z\in\left[\,0.4,\,0.6\right]$ a term $20*rnd$, where
$rnd$ is a random number chosen between $0$ and $1$. The white
noise is different at each time step and at each point of the mesh.
Because the noise is not correlated in space and time, the amplitude
must be taken much higher than in the sinusoidal case (a lot of the
wavelengths generated by this white noise are quickly damped by diffusion
and we wanted to produce a very perturbed flame).

The result of this white noise is shown in Figure \ref{fig:blanc}.
We present also here two flames. The flame on the left is a totally
unstationary flame obtained soon after the first perturbations reach
the tip. The physical parameters are $U\,=\,12$, $\gamma\,=\,0.85$,
and as usual $\varepsilon\,=\,0.1$ $R$$\,=\,$$0.5$.The initial
condition of this calculation was a stationary, unperturbed flame,
with slightly different parameters. Starting from the base of the
flame, the fine grained perturbations directly caused by the white
noise are first seen. Then these perturbations develop because of
the instability, leading to cells where this time, it is difficult
to recognize any hexagon. Then as in the sinusoidal case, the wavelength
of the cells increase because of cell merging. These mergings do not
necessary happen in the same way as the one in Figure \ref{fig:sinus},
where a small cell was swallowed by a large one. More symmetrical
situations are also possible, where the separation between cells disappear
with both cells having approximately the same size. Finally, when
the tip is approached, the perturbations organize in order to produce
a sinuous shape.

This kind of sinuous shape was also observed for two-dimensional flames
\cite{denetbunsen2d}, and an explanation was proposed. But this sinuous
character becomes smaller for typical flames obtained later. We have
an example of one typical flame in the right part of the figure. The
forcing zone, first cells, wavelength increase and finally sinuous
zone can also be seen. It must be admitted however that the flame
is much less sinuous than the flame on the left. The main cause of
this difference is that as the perturbations develop, the flame height
becomes smaller, which can be observed when one compares the two flames
in the figure. We can also insist on the three-dimensional structure
of the sinuous zone to just say that the shape is almost never truly
helical (the author was thinking at the beginning of this work that
it was the natural generalization of the sinuous two-dimensional shape,
but this is apparently not the case). Before closing this section,
one last word, even if we have not performed simulations of turbulent
flames, the two flames of Figure \ref{fig:blanc} can give us some
hope that it can be done with the current three-dimensional model.
But naturally, it would be very expensive, since it is necessary to
either simulate or generate a substitute of turbulent flow. Future
growth in computer power will help, the computations of this article
being done on a relatively standard PC (2.8 Ghz Pentium 4). It must
be noted that the flame lagrangian surface solver used in this article
fits naturally with a lagrangian description of vorticity.

\section{Polyhedral flames \label{sec:Polyhedral}}

In the previous section, the positions of the boundary points were
imposed and the injection velocity was constant. This is a reasonable
approximation of a flame with a very small boundary layer. In this
section however, we are interested in polyhedral flames, which are
well-known experimentally, see for instance \cite{sohrablaw}. A theoretical
(and experimental, but we will be mainly interested in the theoretical
part) article has appeared some time ago \cite{sohrablaw} on this
subject. The polyhedral flames were described by using a modified
version of the Kuramoto-Sivashinsky equation, i.e. making the assumption
that the instability was of thermal-diffusive origin. This type of
model was able to produce polyhedral flames, which is the way cellular
flames look like in the Bunsen burner geometry (the following figures
show these type of flames). However, with the boundary conditions
used in the previous section, we have not been able to produce polyhedral
flames, although we have not made an extensive search, for a lot of
parameters (the calculations for three-dimensional flames are expensive).

So we are faced with a problem. It would not be a nice situation to
conclude that the thermal diffusive instability gives more realistic
results than the hydrodynamic one, as the author, among others, has
claimed in different papers that the hydrodynamic mechanism is the
most realistic. And although not everyone agrees even now, a common
belief seems to be that the thermal diffusive mechanism could only
account for special cases, like lean hydrogen flames. This is probably
why the experiments in \cite{gutmansiva} were performed with hydrogen.

Before concluding that the hydrodynamic model does not contain polyhedral
flames, there is another thing different in \cite{gutmansiva} from
the simulations of the previous section : the boundary conditions.
In \cite{gutmansiva}, the vertical position of the boundary points
was not imposed, and a flame-holder term was added, following Buckmaster
\cite{buckmasterbunsen}, to the Kuramoto-Sivashinsky equation, in
order to modelize attraction of the flame to the burner rim.

So let us just apply here these same conditions. However, we shall
not apply these conditions unmodified, as we have a slight problem.
The vertical coordinate on the boundary is fixed by an equilibrium
between the injection velocity, supposed constant everywhere, and
the flame-holder term (and also the integral terms in equation \ref{mathed:V}).
As a result we have found, and actually it can be seen in the original
article, that the stationary position of the flame is very far from
the flame holder. We propose here, in order to describe flames stabilized
close to the injection zone, to shift this stationary position toward
the flame holder by including the fact that there is a boundary layer,
and that as a result the injection velocity close to the boundary
is much smaller than in the middle of the tube.

We use the following formula for the injection velocity:\begin{equation}
u(r,z)=U-0.9U\exp\left(-\left(a\left(r-R\right)/R\right)^{4}\right)-\beta z\exp\left(-\left(a\left(r-R\right)/R\right)^{4}\right)\label{mathed:u}\end{equation}

where $U$ is the injection velocity at the center of the tube, $a$
and $\beta$ are constant coefficients. . The first exponential term
is there to describe the boundary layer. The $\beta$ term is the
flame holder term used in \cite{gutmansiva}. The exponential variation
with $r$ is the one used in this paper, with coefficient $a$ controlling
the stiffness and width of the boundary layer (proportional to the
inverse of the width). Furthermore, contrary to the previous section,
just on the boundary, the $x$ and $y$ coordinates keep the same
values, but not the $z$ coordinate, which is free to move, being
actually computed as a mean value over the neighbors.

In Figure \ref{fig:polyhedral_U}, we show that these modifications
of the boundary conditions succeed in making polyhedral flames possible
with the Frankel equation. So what is actually important for these
flames is not the mechanism in itself, hydrodynamic or thermal-diffusive,
but what happens close to the boundary, boundary layer, flame holder
term. Actually two flames can be seen in Figure \ref{fig:polyhedral_U}
with parameters $\gamma\,=\,0.85$, $\varepsilon\,=\,0.1$ $R$$=$$0.5$,
$a\,=\,2$, $\beta\,=\,0.4$,$U\,=\,12$, (left) and $U\,=\,6$(right).
These parameters only differ in the injection velocity, showing that,
as also observed in the thermal-diffusive case, this parameter does
not significantly change the cellular structure of the flame. The
eight cells left solution was obtained by taking as initial condition
a flame with no cell, but the eight cells solution on the right was
actually obtained starting from the solution on the left, and apparently
this solution is never completely stationary, as seen in the Figure.

In \cite{gutmansiva}, it was also observed that the parameter $\beta$,
was more important than $U$, and leads to a well-defined transition.
As $\beta$ is increased, the flame is no more cellular. This is also
the case here in the hydrodynamic case, see Figure \ref{fig:polyhedral_beta},
where two flames are shown, the one on the left with $\beta\,=\,0.4$
(same flame as in Figure \ref{fig:polyhedral_U}, left) , and on the
right with $\beta\,=\,20$, all the other parameters being the same.
It can be seen that although this increase in $\beta$ succeeds in
smoothing the flame, the typical values achieving this result are
very high. We have not searched the precise value of the bifurcation,
but for instance the flame is still weakly cellular with $\beta\,=\,8$.
This very high value has the consequence that the flame is closer
from the injection (and above all, closer for a longer distance from
the boundary, which is why the flame height is smaller) and very flat
close to the boundary, which apparently makes this zone stable. On
the contrary, for the small values of $\beta$ used in the other simulations
of this article, the role of the flame holder term seems small.

We have also here another parameter, the width of the boundary layer,
which did not appear in \cite{gutmansiva}. In Figure \ref{fig:polyhedral_blayer},
the effect of this parameter is shown. We take $a\,=\,2$ (left, large
boundary layer), $a\,=\,4$ (right, smaller boundary layer) and $U\,=\,12$,
$\gamma\,=\,0.84$, $\varepsilon\,=\,0.1$ $R$$=$$0.5$, $\beta\,=\,0.4$.
Larger boundary layers correspond to more unstable flames, a prediction
which can tested experimentally (with some caution, i.e. without changing
the flame holder term). This effect could also help explain why it
is observed experimentally that larger injection velocities (and smaller
boundary layers) lead to flames without any cell. It is unclear at
the moment if this effect is sufficient in itself to explain this
stabilization.

This effect suggests the following interpretation of the formation
of polyhedral Bunsen flames. The cellular perturbations form in the
boundary layer. This zone is probably absolutely unstable, and large
boundary layers give more time for the perturbations to develop, which
thus have a larger amplitude at the exit of this zone. Then the perturbations
enter another zone, probably corresponding to a convective instability,
and are convected toward the tip. If the perturbations have not reached
their amplitude at saturation in the boundary layer, they can continue
to amplify during this second phase, leading to cells which appear
at a certain distance from the base of the flame, as in Figure \ref{fig:polyhedral_blayer}
(right), where these cells are very weak. These type of solutions
have indeed been experimentally observed in \cite{gutmansiva} (not
in their simulations however). If one follows this line of reasoning,
the boundary layer width should be compared to typical unstable wavelengths
( most unstable or neutral wavevectors). If this width is much smaller
than the unstable wavelengths, the polyhedral flames should be unlikely
to be observed. This picture is complicated by the effect of the flame
holder term (Figure \ref{fig:polyhedral_beta}) which probably changes
locally the unstable wavelengths, one effect being that flames close
to the injection are also closer from their mirror image, which damps
the Landau instability.

\section{Conclusion\label{sec:Conclusion}}

In this article, we have solved for the first time the Frankel equation
for three-dimensional flames, in the particular case of the Bunsen
burner configuration. This daunting task has been made possible in
a reasonable amount of time by using tools (triangulated surface library
and N-body problem algorithm) made freely available by their authors,
see below the acknowledgments and the references sections. Even with
this help, the treatment of intersections is by no means easy, and
our program far from perfect. We have tried to explain here the algorithms
used to overcome these difficulties. Extensions to other geometries
and to turbulent conditions would be interesting. 

On the Bunsen burner case, we have shown that this equation can at
the same time describe effects absolutely typical of the hydrodynamic
Darrieus-Landau instability (Section \ref{sec:Forcing}) and also
describe polyhedral flames (section \ref{sec:Polyhedral}) in a manner
very close to the existing results obtained with the thermal-diffusive
instability. We have also emphasized the possible importance of the
boundary layer width in this problem. 


\emph{Acknowledgments} : The author would like to thank Stéphane Popinet
for the gts library, Joshua Barnes for his implementation of the Barnes-Hut
algorithm and Geoff Searby for helpful discussions.

\bibliographystyle{unsrt}
\bibliography{./turb2d.reflib}

\section*{List of Figures}

\begin{figure}

\caption{\label{fig:zoom}}

Denet, Phys. Fluids

\includegraphics[%
  scale=0.5]{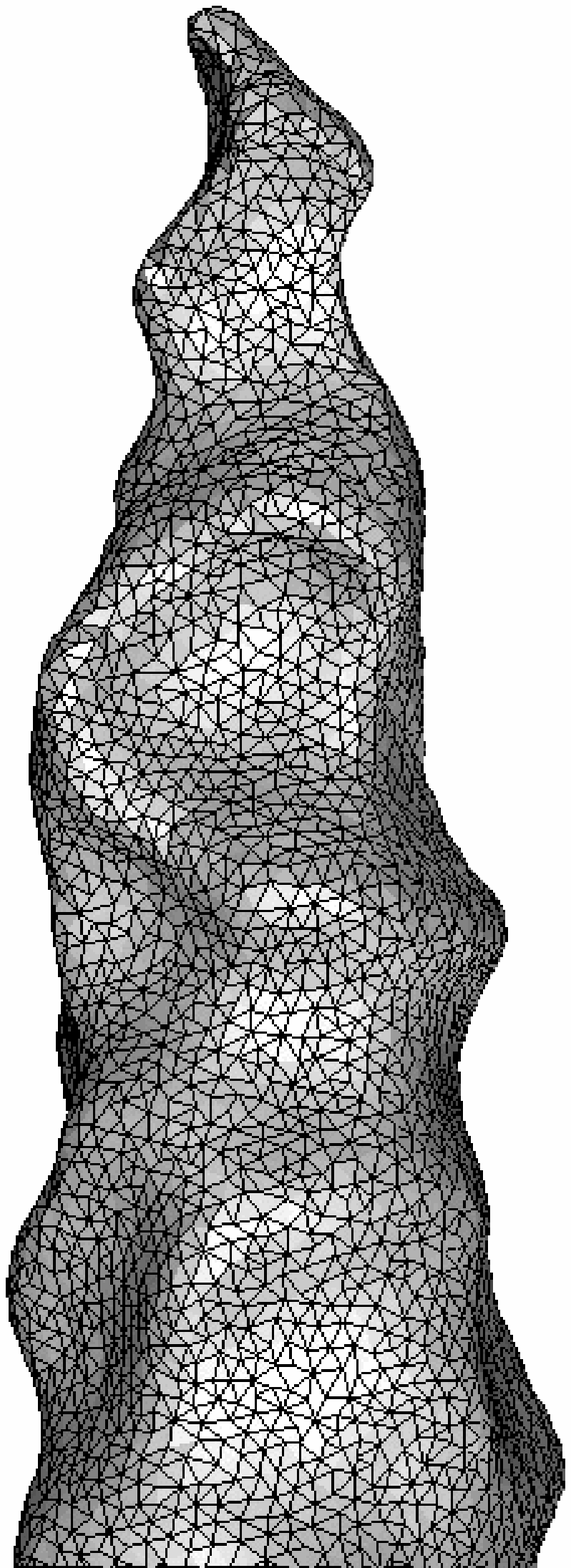}
\end{figure}

\begin{figure}[p]

\caption{\label{fig:blanc} }

Denet, Phys. Fluids

\includegraphics[%
  scale=0.5]{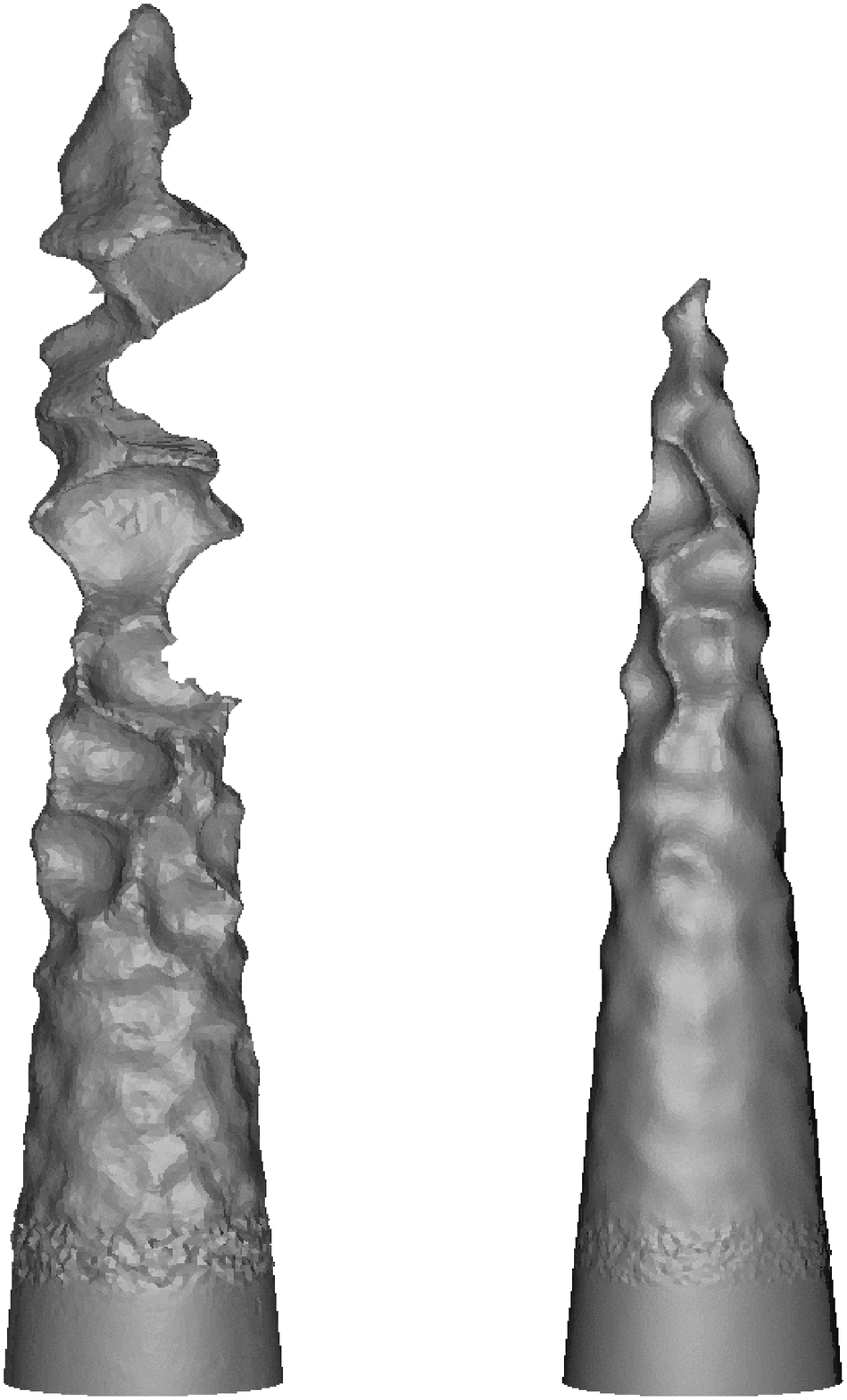}
\end{figure}

\textbf{Figure \ref{fig:zoom}} : zoom on a triangulated surface used
in the simulation (right of Figure \ref{fig:blanc}) 

\begin{figure}[p]

\caption{\label{fig:sinus} }

Denet, Phys. Fluids

\includegraphics[%
  scale=0.5]{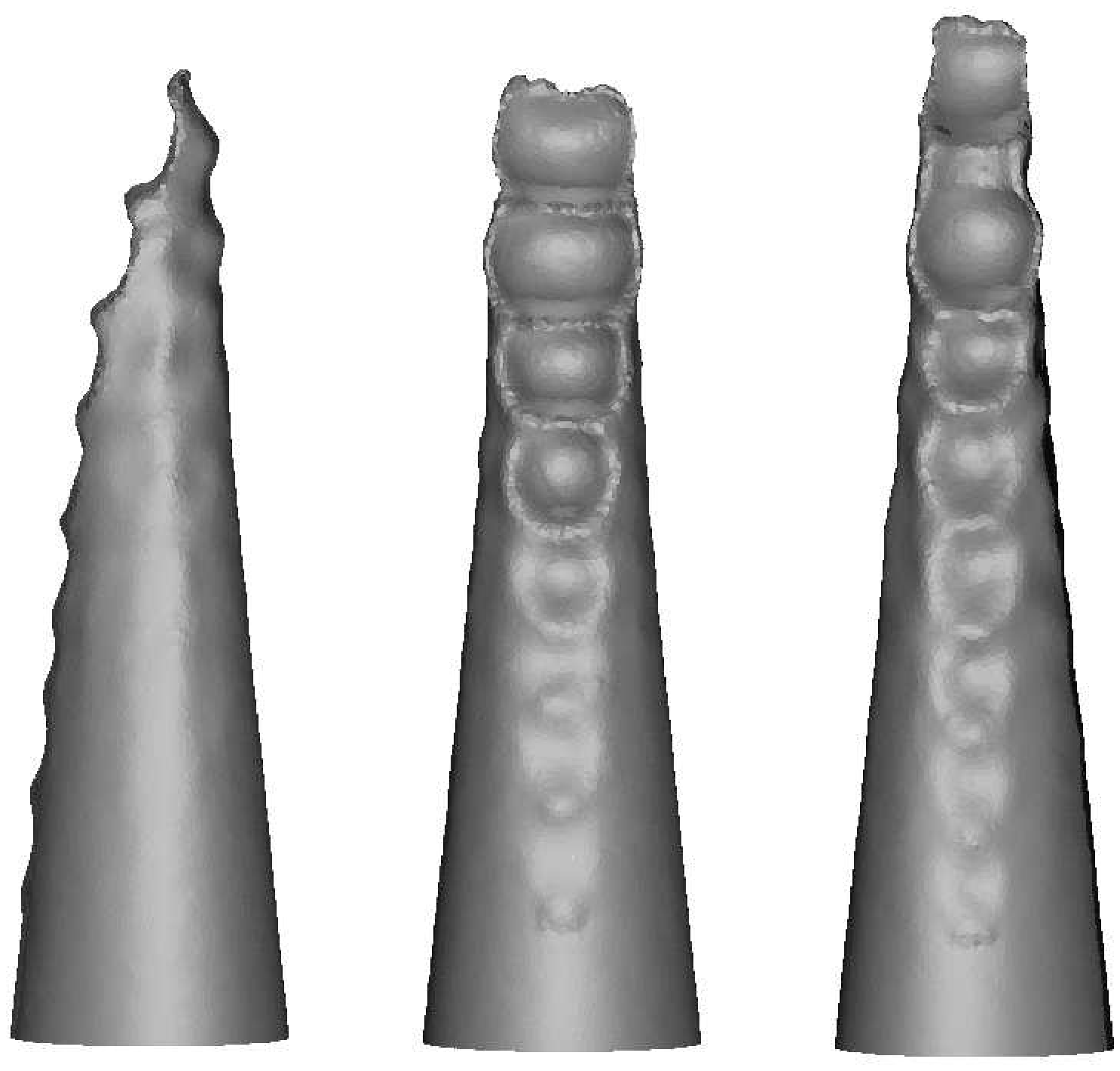}
\end{figure}

\textbf{Figure \ref{fig:sinus}} : Flame submitted to a sinusoidal
forcing. Parameters $U\,=\,12$, $\gamma\,=\,0.85$, $\varepsilon\,=\,0.1$
$R$$=$$0.5$ $\omega\,=\,200$ Left and middle flame: same flame,
seen from two angles, showing the development of the Landau instability
toward the tip. Right: same flame some time later, showing a merging
of two cells close to the tip.

\textbf{Figure \ref{fig:blanc}} : Flame submitted to a white noise
at the base. Left: A flame soon after the instabilities have reached
the tip Right: typical flame some time later. Note the sinuous shape
of the flames close to the tip. parameters $U\,=\,12$, $\gamma\,=\,0.85$,
$\varepsilon\,=\,0.1$ $R$$=$$0.5$

\begin{figure}[p]

\caption{\label{fig:polyhedral_U} }

Denet, Phys. Fluids

\includegraphics[%
  scale=0.7]{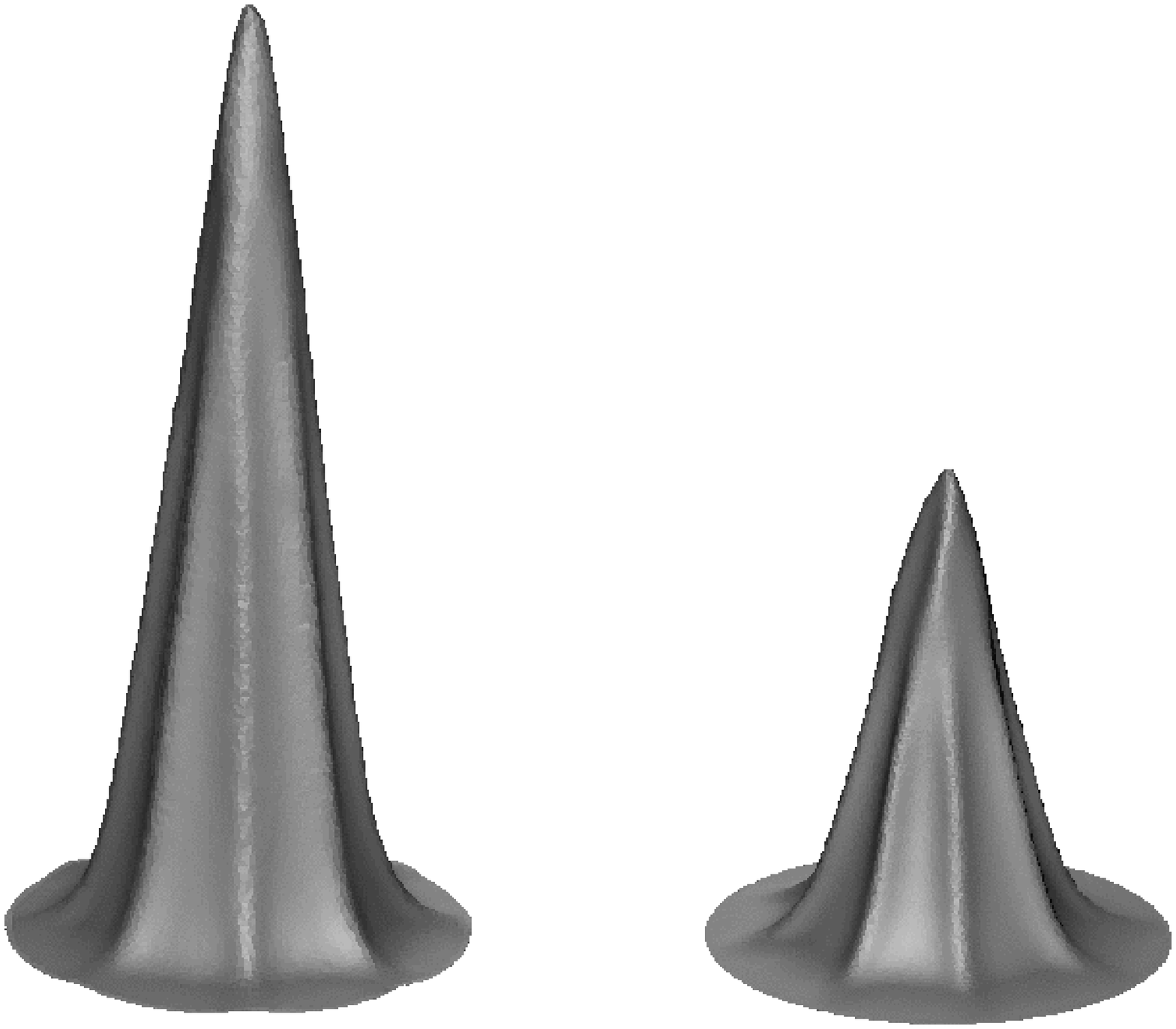}
\end{figure}

\textbf{Figure \ref{fig:polyhedral_U}} : Polyhedral flames obtained
for two different values of the injection velocity parameters $\gamma\,=\,0.85$,
$\varepsilon\,=\,0.1$ $R$$=$$0.5$,$\beta\,=\,0.4$, $a\,=\,2$,
$U\,=\,12$ (left), $U\,=\,6$ (right)

\begin{figure}[p]

\caption{\label{fig:polyhedral_beta} }

Denet, Phys. Fluids

\includegraphics[%
  scale=0.6]{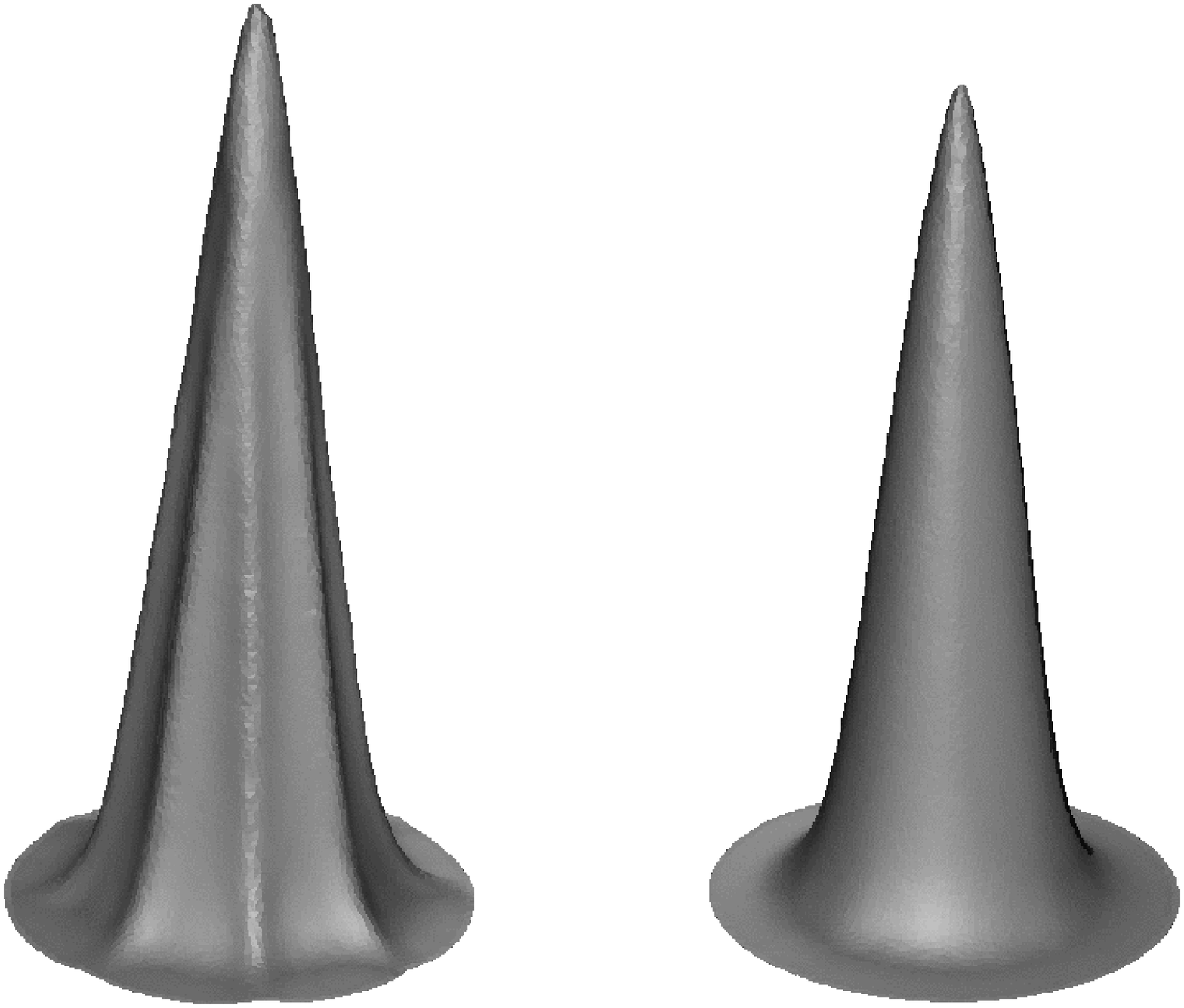}
\end{figure}

\textbf{Figure \ref{fig:polyhedral_beta}}: Flames obtained for two
different values of the flame holder parameter $\beta\,=\,0.4$ and
$20$ other parameters $U\,=\,12$, $\gamma\,=\,0.85$, $\varepsilon\,=\,0.1$
$R$$=$$0.5$, $a\,=\,2$ 

\begin{figure}[p]

\caption{\label{fig:polyhedral_blayer} }

Denet, Phys. Fluids

\includegraphics[%
  scale=0.5]{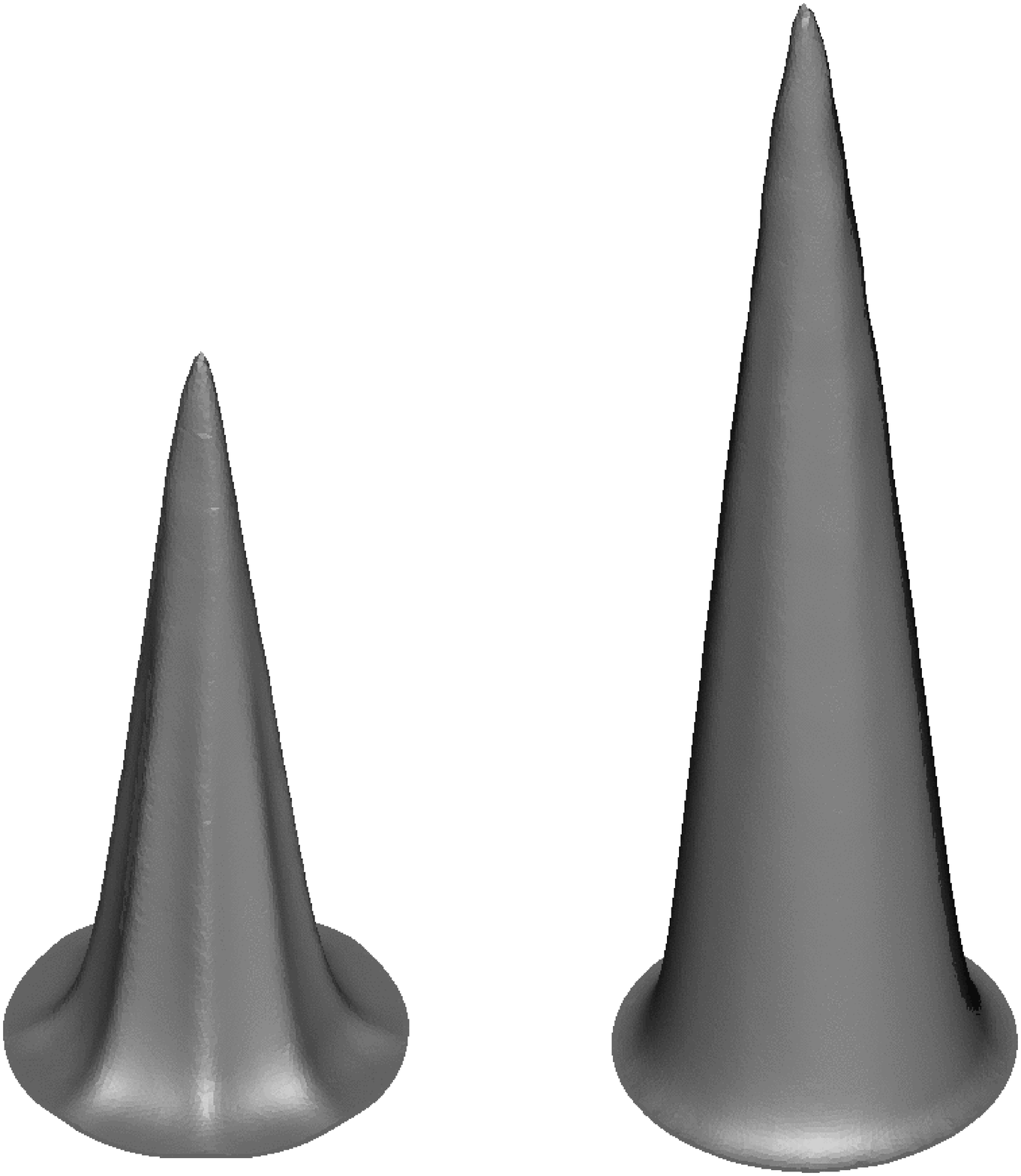}
\end{figure}

\textbf{Figure \ref{fig:polyhedral_blayer}}: Flames obtained for
two different values of the boundary layer thickness, $a\,=\,2$ (left),
$a\,=\,4$ (right) other parameters $U\,=\,12$, $\gamma\,=\,0.84$,
$\varepsilon\,=\,0.1$ $R$$=$$0.5$, $\beta\,=\,0.4$

\end{document}